# Methodical Aspects of Informatization of Physical Education at a Higher Education Institution

*Nurassyl Kerimbayev, Saule Abdykarimova, Saltanat Sharmuhanbet*
*Kazakhstan, Almaty,*
*Abai Kazakh National Pedagogical University*

**Abstract**
The article considers the methodical aspects of informatization of physical education at a higher education institution. Besides, the article discloses the conception of the information educational environment and information technologies oriented at realizing psychological and pedagogical objectives of teaching and educating at a higher education institution. Information technologies in a higher education institution are oriented at realizing psychological and pedagogical objectives of teaching and educating.

The authors of the article show the importance and possibilities of applying information technology media to the system of higher education training of specialists in the sphere of Physics. The authors consider developing the content of methodical training of future Physics teachers to be one of ways of solving the problem of informatization of education.

**Introduction**

The processes of informatization of the modern society enter all the spheres of human activities. This process is especially active in the sphere of education. The introduction of information technologies into different spheres of the education system is getting more scaled and complex. Higher education institutions are becoming leaders of informatization of education, that is the process of providing the whole education system with the theory and practice of developing and applying modern information technologies. Information technologies in a higher education institution are oriented at realizing psychological and pedagogical objectives of teaching and educating.

In the 1990s in domestic and foreign pedagogy, there appears realizing new prospects for informatization of education. Ya.A. Vagramenko, A.V. Khutorsky, N.V. Branovsky stated that information technologies could become the basis of projecting and modeling a new developing environment and educative space, which was called *"information educative space"* and *"information-educational environment"* in a series of research studies[1].

The concept of information-educational environment is described unambiguously and consistently enough. For example, according to L.N. Kechiev, G.P. Putilov and S.R. Tumkovsky, the information-educational environment is a set of computer facilities and methods of their functioning used for realizing teaching activities[2].

The single information-educational environment can mean a program-telecommunication environment based on applying computer hardware and software, and providing students, teachers, parents, university administration bodies and public with single technological facilities. According to the latter definition, the objectives of this environment are the following: giving an information support to the educational process and managing a higher education institution, informing all the members of the educational process about its progress and results, including extracurricular activities.

**Results and Discussion**

The information educational environment includes organizational-methodical facilities, the set of hardware and software for storing, processing and transferring information. These components provide with a ready access to pedagogical information and realize the educational scientific communications, relevant for realizing objectives and learning outcomes of the pedagogical education and developing Pedagogy in today's conditions[3].

Information (Virtual) environment of a higher education institution includes program systems, databases and work techniques supporting the process of managing the organizational activity of a higher education institution[4]. It is necessary to note that these facilities differ from traditional ones in that they are oriented at supporting the process of making decisions by a governing body for the activity of a higher education institution depends mostly on these decisions. Besides, these facilities operate with the integrated set of data, not just with the data divided on departments, services or subsystems of an educational institution.

The single informational (Virtual) space of a higher education institution is the information environment, in which there is a hierarchy of methods for creating information resources and working with them. Any data, information, knowledge, the sources and consumers of which are Bachelor, Master, PhD programs students, teachers, the administration body and the staff of a higher education institution, are considered to be information resources.

One of the most important objectives of informatization of education currently realized in a number of the world community countries, including the Republic of Kazakhstan, is developing people's integral modern scientific ideology adequate to the reality and prospects for development of the global information society, which will be based on wide use of scientific knowledge and new highly effective information technologies[5].

The application of information technologies to teaching Physics helps:
- optimize and modernize the process of teaching;
- use the potential of information technologies, which is inaccessible in the traditional educational process;
- use the potential of multimedia technologies;
- organize various forms of students' activities to gain knowledge independently;
- encourage students' motivation to learn;
- increase their social and professional mobility.

When teaching using modern information and telecommunication technologies, the most important component of the pedagogical process is person-oriented teaching. Such teaching is oriented at teacher–student communication with the use of ICT.

The use of ICT facilities in the system of higher education when training specialists in the sphere of Physics results in beneficiating the pedagogical and organizational activity of an educational institution through the following possibilities:
- Improving methods and techniques of selecting and developing the content of higher professional education in the sphere of Physics;
- Introducing and developing new special courses and disciplines connected with Computer science and IT;
- Development of electronic (digital) educational resources, contents, SCORM-packages;
- Making changes in traditional teaching of the discipline "Physics";
- Increasing effectiveness of teaching–learning due to increasing the level of its individualization and differentiation, and using additional motivational key factors;
- Organization of new forms of communication in the process of teaching–learning and changing the content and character of teacher and learner's activities;
- Improving the mechanism of managing the system of higher education.

Information technologies increase the effectiveness of pedagogical activities. They not only can introduce radical change to understanding the category of the "facility" related to the process of education, but they can strongly influence all the rest components of the system: objectives, content, methods and organizational forms of teaching, educating and personal developing at higher education institutions.

The process of informatization of education makes development of new approaches to applying new information technologies to develop a student's personality, increase the level of his/her activity, develop his/her abilities of developing the strategy of seeking solutions to educational and practical problems and predicting results of realizing the decisions made on the basis of modeling the objects, phenomena, processes under study and their interdependencies a problem of current concern.

However, the process of informatization of higher education institutions management has some essential problems. The existing facilities of informatization applied in higher education institutions are at the stage of development. The costs of purchasing and maintaining equipment are high; there are no specialists on information system development and maintenance, there are no interface, technological and informational links between separate facilities of informatization of education involved in different spheres of activities of higher education institutions.

That is why the important social and pedagogical problem is development of a modern info-communicational culture of future specialists, adequate to the level of ICT development, and these future specialists should satisfy the international indices of competitiveness. Preparedness of future specialists to create new technologies and define a new trajectory of the economic development of the state depends on the extent of this problem realization.

Over the last years, our society has been transformed into an informational society, which makes it necessary to adapt the system of education to the conditions of the new formation. For preparing people for living in an informational society the educational system dictates the need for a new approach to the content, methods and forms of education.

Physics is one of the components of the natural-mathematical subject area, Physics is one of compulsory subjects of the school course. Besides, the course of Physics is to play a key role in developing a scientific picture of the world.

A deep understanding and correct adaptation in the subject-related field of theoretical informatics of physical concepts, their interdependency and inter-conditionality is very important for analyzing processes of informatization of education. The starting premise is association of the concepts: substance, energy, information. "Substance" means the "system", whose components keep the system in balance by mutual transitions from one substance to another.

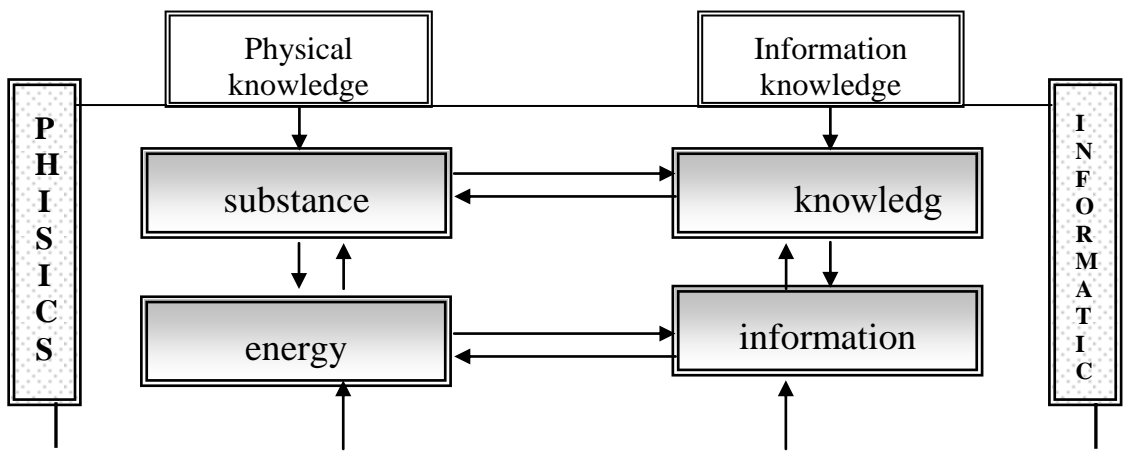

*Picture 1 . Relationship between Informatics (Computer science) concepts and Physics concepts.*

Optimization of the informational process in today's conditions is impossible without using information technologies on the basis of applying methods of projecting modern computer facilities to gain, analyze and store information.

**Conclusion**

The practice of teaching at higher education institutions shows that the education staff in the sphere of Physics has not sufficiently realized the idea of introducing information and telecommunication technologies. In conditions of a plentiful scientific and educational information provided by modern electronic resources requirements for the professional level of Physics teachers increase. However, today there are no specialized educational electronic resources satisfying necessary didactic and methodical requirements. Leading specialists in the sphere of psychology, didactics, content and methods of teaching a specific discipline should be involved in designing the informational resources for using in the teaching process. Most of teachers experience a psychological barrier when they begin to acquire computer devices and apply electronic informational resources to teaching.

Hence, we think that for informatization of education in the sphere of Physics the priority task is adequate training and development of the content of future Physics teachers' methodical training.